\documentclass[a4paper, pra, amsmath, showpacs, preprintnumbers,
superscriptaddress, twocolumn, sort&compress, floatfix]{revtex4}

\usepackage{graphicx, color}

\usepackage{textcomp}
\usepackage[latin1]{inputenc}

\newcommand{\summary}[1]{\mbox{}\marginpar{\raggedright\hspace{0pt}\scriptsize\bfseries[#1]}}
\renewcommand{\summary}[1]{}

\begin{document}

\title{Atom-molecule dark states in a Bose-Einstein condensate}

\author{K. Winkler}
\author{G. Thalhammer}
\author{M. Theis}
\affiliation{Institut f\"ur Experimentalphysik, Universit\"at
  Innsbruck, 
  6020 Innsbruck, Austria}

\author{H. Ritsch}
\affiliation{Institut f\"ur Theoretische Physik, Universit\"at
  Innsbruck, 
  6020 Innsbruck, Austria}

\author{R. Grimm}
\affiliation{Institut f\"ur Experimentalphysik, Universit\"at
  Innsbruck, 
  6020 Innsbruck, Austria}
\affiliation{Institut f\"ur Quantenoptik und Quanteninformation,
\"Osterreichische Akademie der Wissenschaften, 6020 Innsbruck,
Austria}

\author{J. Hecker Denschlag}
\affiliation{Institut f\"ur Experimentalphysik, Universit\"at
  Innsbruck, 
  6020 Innsbruck, Austria}

\date{\today}


\begin{abstract}
We have created a dark quantum superposition state of
  a Rb Bose-Einstein condensate (BEC) and a degenerate gas of Rb$_2$
  ground state molecules in a specific
  ro-vibrational state using two-color photoassociation.
As a signature for the decoupling of this coherent atom-molecule
gas from the light field we observe a striking suppression of
photoassociation loss. In our experiment the maximal molecule
population in the dark state is limited to about 100 Rb$_2$
molecules due to laser induced decay. The experimental findings
can be well described by a simple three mode model.
\end{abstract}

\maketitle

\summary{cold molecules}%

 The phenomenon of
coherent dark states is well known in quantum optics and is based
on a superposition of long-lived system eigenstates which
decouples from the light field. Since their discovery \cite{Ari76}
dark states have found numerous applications. Prominent examples
are electromagnetically induced transparency and lasing without
inversion \cite{Har97}, sub-recoil laser cooling \cite{Asp88}, and
ultra-sensitive magnetometers \cite{Sta02}. A particular
application is the coherent transfer of population between two
long-lived states by a stimulated Raman adiabatic passage (STIRAP)
\cite{Bergmann}.

In the emerging field of ultracold molecules, the conversion of
atomic into molecular BECs is a central issue. A series of recent
experiments on the creation of molecular quantum gases rely on the
application of Feshbach resonances \cite{molecules}. This coupling
mechanism, however, is restricted to the creation of molecules in
the highest ro-vibrational level and is only practicable for a
limited number of systems. As a  more general method a stimulated
optical Raman transition can directly produce deeply bound
molecules as demonstrated a few years ago \cite{Wynar,Tol01}.
STIRAP was proposed as a promising way for a fast, efficient and
robust process to convert a BEC of atoms into a molecular
condensate \cite{Vardi, Jul98, Mackie, Hop01, Drummond, Dam03}.
The central prerequisite for this kind of STIRAP is a dark
superposition state of a BEC of atoms and a BEC of molecules.

In this Letter, we report the observation of such a collective
multi-particle dark state in which atoms in a BEC are pairwise
coupled coherently to ground state molecules.
\begin{figure}
  \includegraphics{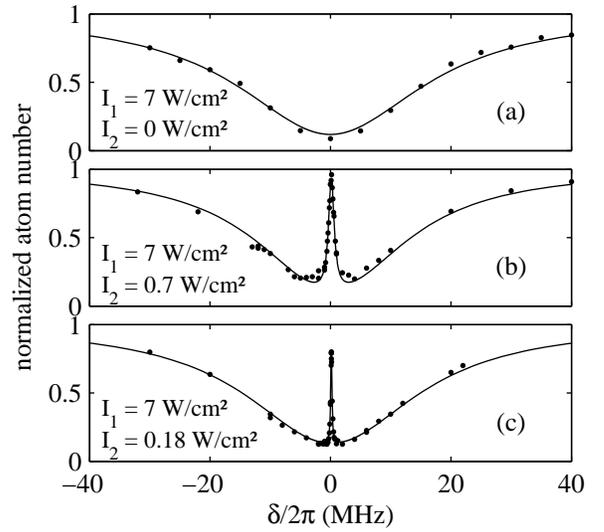}
  \caption{Dark resonances in two-color photoassociation. (a) Atomic loss signal
  in one-color photoassociation as a function of the laser detuning
   from the electronically excited molecular line.
    (b), (c) When we apply a
    second laser (fixed frequency)
    which resonantly ($\Delta$ = 0) couples the excited molecular state to a long
    lived molecular ground state, the losses are strongly suppressed
    at ${\delta}=0$.  Depending on
    the intensity of laser 2, this dark resonance can get very
    narrow. The atom life time on the dark resonance in (b) is
     140 ms whereas in (a) atoms have an initial decay time of about 2 ms.
     Intensities of laser 1 (I$_1$) and 2 (I$_2$) are as indicated. }
  \label{fig:autler-narrow}
\end{figure}
This dark atom-molecule BEC shows up in a striking suppression of
photoassociative loss, as illustrated by the spectra in
Fig.~\ref{fig:autler-narrow}. In one-color photoassociation, the
excitation of a molecular transition produces a resonant loss
feature that reflects the optical transition linewidth, see
Fig.~\ref{fig:autler-narrow}(a). The presence of a second laser
field coupling the electronically excited molecular state to a
long-lived ground-state level can drastically reduce this loss, as
shown in Fig.~\ref{fig:autler-narrow}(b) and (c). In (b), for
example, we observe a striking loss suppression by about a factor
of 70 on resonance.

Already the mere observation of an atom-molecule dark resonance in
a BEC proves that a coherent, quantum degenerate gas of molecules
has been formed. This follows from the facts that 1) the dark
state is by definition a coherent superposition of atoms and
molecules and 2) the atomic BEC is a coherent matter-wave. In this
fully coherent situation, the molecular fraction itself must be
quantum degenerate with a phase-space density corresponding to the
number of molecules. The very narrow resonance lines indicate the
high resolution of our measurements and the potential sensitivity
of the dark state as an analysis tool. Using a BEC allows direct
interpretation and clear understanding of our data without
ambiguity. Thermal averaging of signal features plays no role in
contrast to previous measurements in thermal gases \cite{Sch03,
Tol01, Lis02}.

\begin{figure}
  \input{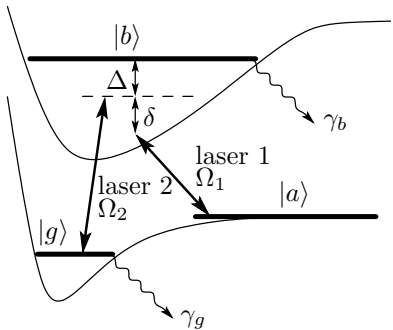}
  \caption{Level scheme.
 $\Delta$ and $\delta$ denote the detunings. $\Omega_1$ and $\Omega_2$
 are the Rabi frequencies. The excited molecular
    state $|b\rangle$ spontaneously decays with a rate $\gamma_b$
  to levels outside this scheme.  The molecular
  state $|g\rangle$ is attributed a  decay rate $\gamma_g$
   which phenomenologically takes into account
   losses through inelastic collisions and laser induced
   dissociation, e.g. when laser 1 couples $|g\rangle$ to the
  unstable state $|b\rangle$.
     In all our
   measurements laser~1 is scanned (varying ${\delta}$) while
    laser~2 is held fixed at a particular detuning ${\Delta}$.}
  \label{fig:raman-scheme}
\end{figure}

\summary{exp: Rb, magn. trap}%
The starting point of our measurements is a BEC of $4\times 10^5$
$^{87}$Rb atoms in the spin state $| F = 1, m_F = -1 \rangle$
\cite{Thalhammer}. In the level scheme of
Fig.~\ref{fig:raman-scheme} the atomic BEC state is represented by
$|a\rangle$. Laser 1 couples this state to the excited molecular
state $|b\rangle$. Laser 2 couples $|b\rangle$ to the molecular
ground state $|g\rangle$. We choose level $|b\rangle$ to be the
electronically excited molecular state $|0_g^-,\nu = 1, J = 2
\rangle$ located $26.8\,\text{cm}^{-1}$ below the $S_{1/2} +
P_{3/2}$ dissociation asymptote \cite{Thalhammer}. For level
$|g\rangle$ we choose the second to last bound state in the ground
state potential. It has a binding energy of $E_\text{b}/h =
636\,\text{MHz} $ \cite{Wynar}.
 $|a\rangle$, $|b\rangle$ and $|g\rangle$ form the lambda-system
for the atom-molecule dark states.

We illuminate the trapped condensate for typically 10\,ms with two
phase-locked laser beams in a Raman configuration as shown in
Fig.~\ref{fig:raman-scheme}. Both laser beams are derived either
from a single diode laser or, for higher optical powers, from a
Ti:Sapphire laser.  The frequency difference between the two beams
is created with an acousto-optical modulator at a center frequency
of about 320\,MHz in a double-pass configuration. This allows
precise control of the beams' relative frequency difference over
several tens of MHz. Both beams propagate collinearly and are
aligned along the weak axis of the trap. They have a waist of
about 100\,\textmu{}m, and their linear polarization is
perpendicular to the magnetic bias field of the trap.
The diode laser and the Ti:Sapphire laser both have line widths of
less than 1\,MHz. They are offset locked relative to the
$D_2$-line of atomic rubidium with the help of a scanning optical
cavity. This yields an  absolute frequency stability of better
than 10\,MHz.
\summary{Fig 3: high power spectra}%
\begin{figure}
  \includegraphics{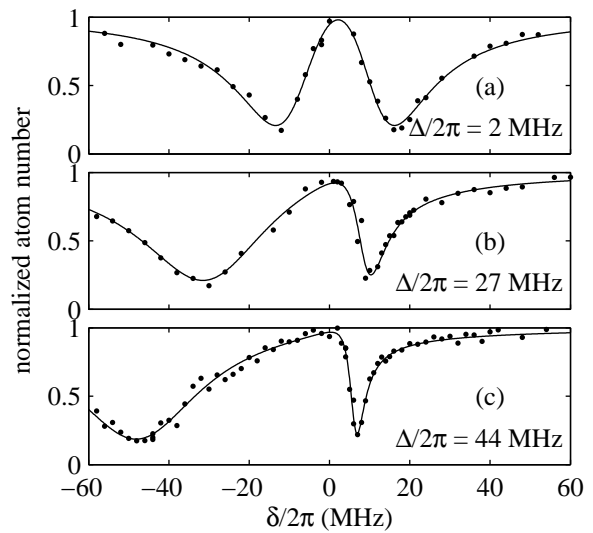}
  \caption{Two-color photoassociation spectra for various detunings
  $\Delta$ at a large intensity  $I{_2} = 20$\,W/cm$^2$.  Here $I{_1} =
    80$\,W/cm$^2$. The solid lines are fit
    curves based on our theoretical model.}
    \label{fig:autler}
\end{figure}

We are able to describe all our spectra  with a relatively simple
three mode model. Although the atom-molecule dark states are
intrinsically complicated and entangled, in a first approximation
the atoms and molecules can be represented as coherent matter
fields  \cite{Vardi, Jul98, Mackie, Hop01, Drummond, Dam03}. Using
the notation of M.~Mackie \emph{et al.} \cite{Mackie} we obtain a
set of differential equations for the normalized field amplitudes
$a$, $b$, and $g$ of the BEC state, the excited molecular and
ground state, respectively:
\begin{equation}
  \begin{aligned}
    i \dot{a} &= -\Omega_1 a^* b,\\
    i \dot{b} &= \left[(\Delta+\delta)-i\gamma_b/2 \right ] b
    - \tfrac{1}{2}(\Omega_1 aa + \Omega_2 g),\\
    i \dot{g} &= (\delta - i\gamma_{g}/2)g
    - \tfrac{1}{2}\Omega_2 b. \label{equ:formel}
  \end{aligned}
\end{equation}

We refer to $\Omega_{1}$ as the free-bound Rabi frequency (see
Fig.~\ref{fig:raman-scheme}). It scales with intensity $I_1$ of
laser 1 and initial atom density $\rho$ as $\Omega_1 \propto
\sqrt{I_1}\sqrt{\rho}$. The bound-bound Rabi frequency $\Omega_2
\propto \sqrt{I_2}$  only depends on the intensity $I_2$ of laser
2. The detunings $\Delta$ and $\delta$ are defined as depicted in
Fig.~\ref{fig:raman-scheme}. $\gamma_{b}$ and $\gamma_{g}$ denote
the effective decay rates of state $|b\rangle$ and $|g\rangle$
(for details see Fig.~\ref{fig:raman-scheme}). $|a|^2,|b|^2$, and
$ |g|^2$ give the ratio between the respective atom (molecule)
number and the initial atom number. In the absence of losses, i.e.
$\gamma_{b} = \gamma_{g} = 0$, particle numbers are conserved
globally,  $|a|^2 + 2 |b|^2 + 2 |g|^2 = 1$. Unlike the previous
theoretical treatments \cite{Vardi, Jul98, Mackie, Hop01,
Drummond, Dam03} where the decay rate $\gamma_{g}$ was basically
neglected, we find that $\gamma_g$ is relatively large and
intensity dependent, $\gamma_g = \gamma_g (I_1)$. In our simple
model we do not include atomic continuum states other than the BEC
state. We neglect inhomogeneity effects due to the trapping
potentials and finite size laser beams. Energy shifts caused by
the mean-field interaction of atoms and molecules are small and
 neglected.

In order to determine the parameters of our model and to check it
for consistency, we performed measurements in a broad parameter
range of intensities and detunings.  Fits to the photoassociation
curves determine all unknown parameters of the system such as
${\Omega_{1}}$, ${\Omega_{2}}$, $\gamma_{b}$ and $\gamma_{g}$.
Figure~\ref{fig:autler} shows photoassociation spectra for a
relatively high laser power $I_2 =$ 20 W/cm$^2$ and various
detunings $\Delta$.  For a small detuning ${\Delta}$
(Fig.~\ref{fig:autler}~(a)) the dark resonance line from
Fig.~\ref{fig:autler-narrow} has broadened considerably. This
spectrum can also be viewed as two absorption lines resulting from
a strong Autler-Townes splitting which was also  observed
 in  thermal gases \cite{Tol01,Sch03}. From the 30 MHz separating the
two resonance dips, the magnitude of the Rabi frequency
${\Omega_{2}}$ can be directly determined. For a larger detuning
${\Delta}$, the resulting spectrum becomes asymmetric and turns
into a narrow and a broad dip, see Fig.~\ref{fig:autler}~(b) and
(c). The narrow loss feature is related to the two-photon Raman
transition while the broad dip is due to the one-photon transition
$|a\rangle \rightarrow |b\rangle$.
 Note that similar to Fig.~\ref{fig:autler-narrow}, losses are
 suppressed at ${\delta}=0$.

\begin{figure}
  \includegraphics{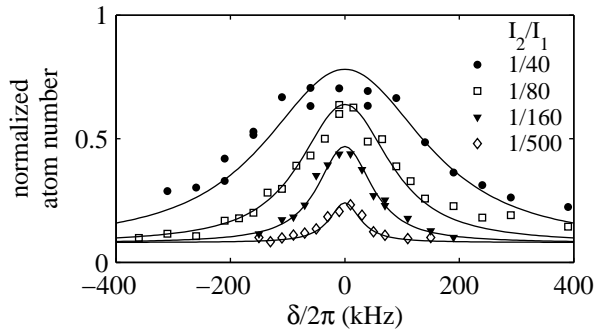}
  \caption{Dark resonances (blow-up of the central part of
  Fig.~\ref{fig:autler-narrow}~(c) and similar
  curves) for
    different intensity ratios $I_2/I_1$  (see legend) at a fixed intensity
    $I_1 = 7\,\text{W}/\text{cm}^2$.
    The solid curves are calculations based on our theoretical
    model. $\Delta = 0$.
    }
    \label{fig:lines}
\end{figure}

\begin{figure}
  \includegraphics{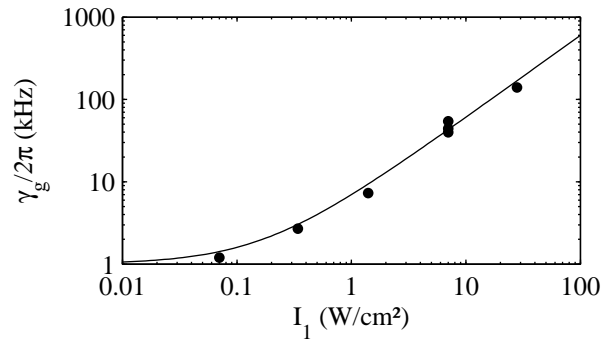}
  \caption{Dependence of the decay rate $\gamma_g$ of the ground state molecules
     on the laser intensity $I_1$, measured with an
    intensity ratio $I_2/I_1 = 1/40$. The solid curve
    is given by $\gamma_g
     =  \ 2\pi \times 6\,\text{kHz}/(\text{W\,cm}^{-2}) I_1 +
 2 \pi \times 1\,\text{kHz}. $
 }
    \label{fig:dependence}
\end{figure}

Figure~\ref{fig:lines} shows the dark resonances in the low power
limit where $I_1$ is held constant and $I_2$ is lowered in 4
steps. The dark state transforms more and more into a grey state,
because losses become more dominant due to a nonzero decay rate
$\gamma_g$. The height of the dark resonance decreases when the
pumping rate ${\Omega_{2}^2/\gamma_{b}}$ comes in the range of the
decay rate of the molecular ground state $\gamma_{g}$. This allows
 for a convenient determination of $\gamma_g$.
From Fig.~\ref{fig:lines} it is also clear that the width of the
dark  resonance decreases with $\Omega_2$. For $\Omega_2 \ll
\gamma_b$ the width is given by $\Omega_2^2/\gamma_b + \gamma_g$,
corresponding to power broadening and the effective ground state
relaxation.
The following set of parameters describes all our measurements
quite accurately and was used in particular for the calculated
solid lines in Fig.~\ref{fig:lines}: $\Omega_1/(\sqrt{I_1}
\sqrt{\rho/\rho_0}) =
2\pi\times8\,\text{kHz}/(\text{W\,cm}^{-2})^{1/2}$ %
at a peak density of $\rho_0 = 2 \times10^{-14}\,\text{cm}^{-3}$, %
$\Omega_2 / \sqrt{I_2} =
2\pi\times7\,\text{MHz}/(\text{W\,cm}^{-2})^{1/2}$, and %
$\gamma_b = 2\pi\times13\,\text{MHz}$.  We find that the decay
rate $\gamma_g$ of the ground state molecular level increases with
the intensity $I_1$ of laser 1 as shown in
Fig.~\ref{fig:dependence}. A dependence of $\gamma_g$ on $I_2$ was
negligible in our experiments where typically $I_2/I_1 \sim   1/5
\dots 1/500$.
 We model the behavior of $\gamma_g$ as
$\gamma_g =   2\pi \times 6\,\text{kHz}/(\text{W\,cm}^{-2}) I_1 +
\gamma_\text{bg}$, the sum of  a light-induced decay rate
proportional to $I_1$ and background decay rate $\gamma_\text{bg}$
due to inelastic collisions in the absence of light.  From
measurements at low intensities we can estimate an upper value for
the background decay rate of about $\gamma_\text{bg} \approx
2\pi\times1\,\text{kHz}$ for $\rho_0  =  2\times
10^{14}\,\text{cm}^{-3}$. This value for $\gamma_\text{bg}$ is
consistent with previous experimental
results for $^{87}$Rb at similar atom densities \cite{Wynar}. %
The increase of $\gamma_g$ with $I_1$ is due to several
imperfections which break the ideal 3-level lambda system. Laser~1
also couples the molecular ground state $|g\rangle$ to the
short-lived excited molecular state $|b\rangle$, which leads to an
incoherent loss of the molecules due to spontaneous decay. Due to
the rather small frequency difference ($\approx 2\pi \times 636$
MHz) of the two Raman lasers and the strong bound-bound
transition, this cannot be neglected. In addition, only 290 MHz
 below  level $|b\rangle$ exists another
 excited molecular state  $|0_g^-,\nu = 1, J = 0 \rangle$ which
represents an additional loss channel \cite{Thalhammer}. These two
contributions explain about one third  of our observed losses.
Furthermore, losses can also stem from a photodissociation
transition which couples ground state molecules  directly to the
continuum above the $S_{1/2} + P_{1/2}$  dissociation asymptote.

 Having determined the  parameters we can use  model
 (\ref{equ:formel})
 to calculate the fraction of ground state molecules $|g|^2$.
 For the measurements presented in Fig.~\ref{fig:lines} we have a
peak molecular fraction of $2\times10^{-4}$ corresponding to about
100 molecules  (at $\delta = 0 $ and $ I_2 / I_1 = 1 / 500$). For
comparison, for $ I_2 / I_1 = 1 / 40$ the molecule number is only
about 25 at $\delta = 0 $. It is interesting to note how few
molecules are needed to stabilize almost a million atoms against
photoassociation. This large asymmetry of the particle numbers
reflects the different coupling strengths of the free-bound and
bound-bound transitions.
Naturally the question arises how  the experimental parameters
should be chosen to optimize the number of molecules. This is
non-trivial due to the finite decay rate $\gamma_g$. With model
(\ref{equ:formel}) we have numerically mapped out molecule numbers
as a function of time, detuning and laser intensities, starting
out with a pure atomic BEC and simply switching on the lasers. In
general within a few $\mu$s of evolution, the dark state is
formed. This involves only negligible losses of atoms since the
dark state is very close to our initial BEC state.
\begin{figure}
  \includegraphics{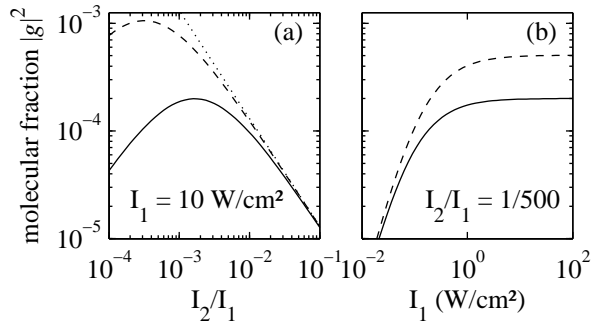}
  \caption{(a) Maximum molecular fraction as function of the intensity ratio
    $I_2/I_1$ at a fixed intensity $I_1 = 10
    \text{W}/\text{cm}^2$ and (b) as a function of  intensity
    $I_1 $ at a fixed intensity ratio $I_2/I_1 =  1/500$.
    The solid lines show the molecule fraction for the measured decay
    rate $\gamma_g = 2\pi\times 6\,\text{kHz}/(\text{W\,cm}^{-2}) I_1
    + \gamma_\text{bg}$.  The dashed lines show the molecule fraction
    assuming a lower decay rate $\gamma_g = 2\pi\times
    1\,\text{kHz}/(\text{W\,cm}^{-2}) I_1 + \gamma_\text{bg}$. The dotted line
    corresponds to $\gamma_g = 0$. The
    calculations are based on Eqs.~(\ref{equ:formel}) with $\Delta = 0$.}
    \label{fig:molecules}
\end{figure}
 The maximum number of molecules of every evolution is then
determined. We find that we can optimize the molecular production
by working at $\Delta = 0$ although other values for $\Delta$ can
be used.
 For $\Delta=0$ the maximum number of molecules correspond
to $\delta = 0$, hence both lasers are on resonance.
Figure~\ref{fig:molecules} shows the molecular fraction as a
function of the laser intensities. In Fig.~\ref{fig:molecules}(a),
as $I_2 / I_1$ is lowered from high values, the molecule fraction
initially grows and follows a straight dotted line which coincides
with the ideal route for STIRAP. Following this route would lead
to a full conversion of atoms into molecules in the absence of
loss $\gamma_g$. This can be seen from Eqs.~(\ref{equ:formel})
when setting $  \dot{b} = 0$ and $b =0 $ such that $ |g|^2 =
\Omega_1^2 /\Omega_2^2 |a|^4 $. For finite $\gamma_g$, however,
the molecular fraction curve rolls over for some value of $I_2 /
I_1$, when the molecule loss rate is larger than its production
rate. A smaller $\gamma_g$ would lead to a larger number of
molecules (dashed line). The finite $\gamma_g$ in our experiments
leads to a maximum molecule number at $I_2/I_1 \sim 1/500$, a
ratio which we also used in our measurements (see
Fig.~\ref{fig:lines}, open diamonds). For this optimum value
 the dependence of the molecular fraction on $I_1$
is shown in Fig.~\ref{fig:molecules}(b). Here it becomes clear
that the laser intensities have to be kept above a certain
threshold so that losses are not dominated by the background decay
rate $\gamma_\text{bg}$ of the molecular state.

To summarize, we have created a novel multi-particle dark state
where an optical Raman transition coherently couples an atomic Rb
BEC of about 4$\times$10$^5$ atoms to a quantum degenerate gas of
up to 100 Rb$_2$ ground state molecules. Our investigations can be
extended in a straight forward manner to create and study BECs of
arbitrarily deeply bound molecules and coherent atom/molecule
mixtures. The dark resonance has proven itself as a useful tool to
analyze the atom-molecule system and to optimize the optical
conversion of atomic to molecular BECs. An increase of the number
of molecules by several orders of magnitude
 should be possible by choosing better suited
ground and excited molecular states for the free-bound Raman
transition.

We appreciate the help of George Ruff and Michael Hellwig at an
early stage of the experiment. We thank Paul Julienne, Eite
Tiesinga, Peter Drummond and Karen Kheruntsyan for valuable
discussions. This work was supported by the Austrian Science Fund
(FWF) within SFB 15 (project parts 12 and 17) and the European
Union in the frame of the Cold Molecules TMR Network under
contract No.~HPRN-CT-2002-00290.

\end{document}